# INJECTION BUCKET JITTER COMPENSATION USING PHASE LOCK SYSTEM AT FERMILAB BOOSTER*

K. Seiya[†], C. Drennan, W. Pellico, S. Chaurize, FNAL, Batavia, U.S.A.

*Abstract*

The extraction bucket position in the Fermilab Booster is controlled with a cogging process that involves the comparison of the Booster rf count and the Recycler Ring revolution marker. A one rf bucket jitter in the extraction bucket position results from the variability of the process that phase matches the Booster to the Recycler. However, the new slow phase lock process used to lock the frequency and phase of the Booster rf to the Recycler rf has been made digital and programmable and has been modified to correct the extraction notch position. The beam loss at the Recycler injection has been reduced by 20%. Beam studies and the phase lock system will be discussed in this paper.

## FERMILAB 700KW OPERATION

Fermilab has provided a 700kW proton beam to the NOvA neutrino experiment since January 2017. Improvements on the reduction of the beam loss in the Linac, Booster, Main Injector and Recycler made great strides toward reaching this goal under the Proton Improvement Plan project. [1] The beam loss thresholds, in each machine, set radiation levels for the sake of the equipment, the environment and the personnel that must maintain the beamline components. As beam intensities increase, more sources of beam loss in the machines must be addressed to reach the higher operating intensities. A beam loss had been occurring at the Recycler injection due to the synchronization errors between the Booster and Recycler. This had been an issue since the Booster first implemented a beam extraction synchronization system, called Cogging, in 2002. [2]

## BEAM SYNCHRONIZATION FROM BOOSTER TO THE RECYCLER

The Booster, a resonant circuit synchrotron, accelerates a proton beam from 400MeV to 8GeV and sends the batches of protons to the Recycler at a rate of 15Hz. The rf frequency changes from 37.9MHz to 52.8MHz. One Booster batch has 81 filled buckets and 3 empty buckets. (Figure 1) The empty buckets, referred to as the Notch, are positioned to be at the Booster extraction point precisely timed to the rising extraction kicker voltage, and then at the Recycler injection kicker as its voltage is rising. The rise-times of kickers are about 40 nsec which requires a notch of more than two 52.8 MHz rf buckets.

When the position of the Notch is mistimed, beam loss can occur at the Booster extraction. If the Booster extraction timing is adjusted to correct for the mistimed Notch, the Booster batch will land a bucket position off in the Recycler and then a beam loss occurs at the Recycler injection point. Since two asynchronous systems, Cogging and Phase Lock, both effect the Notch position, a +/- 1 bucket jitter is common. The following sections will explain Notch creation at Booster injection, Recycler injection, Cogging and Phase Lock and how the jitter results.

### Notch Creation

The beam is injected from the LINAC to the Booster at 400MeV, with a 200MHz structure and is captured adiabatically with a 37.9MHz rf. The harmonic number is 84. As soon as the beam is captured and bunched into rf buckets, three out of the 84 bunches are kicked out of the Booster, into an absorber. The three empty buckets are the Notch. The Notch creation occurs early in the Booster cycle to minimize the radiation produced from the kicked bunches.

### Recycler Injection

The Recycler is a fixed energy synchrotron using permanent magnets and has a harmonic number of 588. A total of 12 batches are sent to the Recycler. The Recycler has space for 7 Booster batches. At the beginning of a cycle, the Recycler is injected with 6 Booster batches end to end. Then 6 more batches are injected using the 7th position and slip stacked onto the previous 6 batches. [3] At the end of the cycle these 12 batches are transferred to the Main Injector.

### Cogging

The new Magnetic Cogging system [4] was implemented in the Booster in 2014 to control the position of the Notch through the cycle, and to synchronize the Notch with the Booster extraction and the Recycler injection. The position of the Notch is changed by manipulating existing dipole correctors in the Booster. The control logic is implemented within a programmable VXI module using the Booster rf and the Recycler rf as clocks.

While the Booster is accelerating beam, the Recycler stays at a fixed frequency of 52.8MHz and sends timing pulse called the OAA markers. The OAA marker indicates the Recycler bucket position for the next Booster batch injection. The OAA marker is synchronized to the Recycler rf and sent out every Recycler revolution.

The Cogging feedback system keeps the number of Booster RF counts between the OAA markers the same as it was recorded to be on the reference cycle. The reference cycle is the Booster cycle that delivers the first batch of beam to the Recycler.

___________________________________________

* Work supported by Fermilab Research Alliance, LLC under Contract No. DE-AC02-07CH11359 with the United States Department of Energy.
† kiyomi@fnal.gov

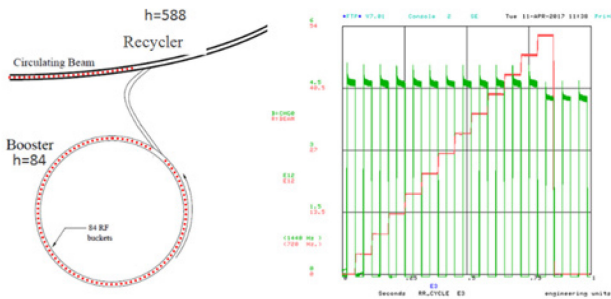

Figure 1: Illustration of beam delivery to the Recycler and beam intensities in the Booster (green) and Recycler(red).

*Phase Lock*

The Booster rf must be at the same frequency as the Recycler rf at extraction. Additionally, the phase relationship of the Booster rf to the Recycler rf needs to be at a consistent phase offset at extraction. This is accomplished by switching from the acceleration phase lock of the Booster rf and the beam, to a "slow" phase lock of the Booster rf and the Recycler rf. [5] This occurs approximately 3 msec before extraction. This slow phase lock is accomplished by dividing the frequency of both the Booster and Recycler rf signals by 32 using counters. A feedback signal is computed to adjust the frequency of the Booster rf source so that the divide-by-32 phase difference follows a reference curve, settling just before extraction.

Figure 2 shows the main signals of interest in the slow phase lock process. The green trace is the phase difference between the un-divided Booster and Recycler rf. The triangular wave shape indicates that they are at different frequencies. The waveform changes from peak to peak as the phase between them rotates from 0 degrees to 180 degrees on to 360 degrees, back to 0 degrees. The number of periods of the triangle wave represents how many times the Booster rf periods slide by the Recycler rf periods before settling with a flat slope indicating the frequency are then equal. When the final value of this signal is consistent from cycle to cycle, it indicates that the extraction phase difference of the rf signals are consistent from cycle to cycle.

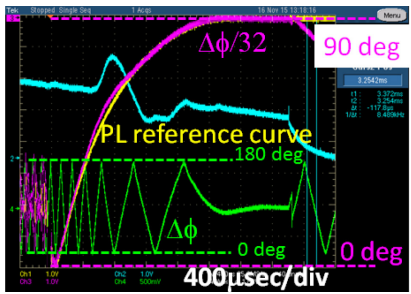

Figure 2: Illustration of the slow phase lock signals.

The magenta trace in the figure is the divide-by-32 phase difference. Using a feedback signal to the Booster rf source, the cyan trace, the divide-by-32 phase difference is made to follow the yellow reference curve to a match of the divide-by-32 frequencies and a final divide-by-32 phase offset.

*Jitter on the Recycler Injection Bucket Position*

Jitter in positioning the Recycler injection bucket arises from two effects. One is counting the Booster rf buckets with respect to a signal that is not synchronous to this rf, the OAA marker. The second is that the phase lock process, which takes control away from the cogging process, has its own criteria for when it must start. This causes the cogging process to be terminated at different times in its progression. Typically, the position of the injection bucket relative to the OAA has a plus or minus one bucket jitter, with respect to the reference cycle, when the phase lock control is turned on.

Figure 3 shows the Wall current monitor (WCM) signal at the Recycler injection for the first 6 batches with a scale of 10 μsec per trace.

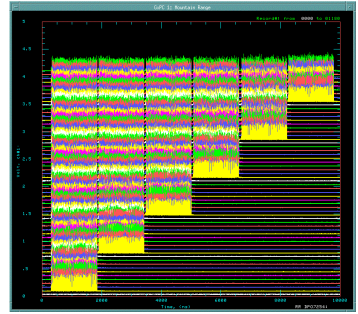

Figure 3: Mountain Range plots of Recycler beam for the first 6 batch injections.

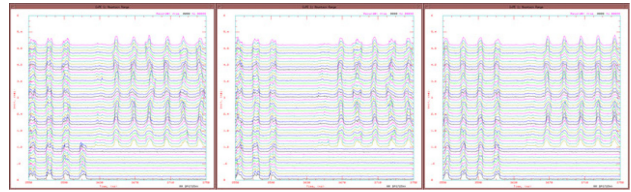

Figure 4: Mountain Range plots which show the WCM signal at the 3rd batch injection in the Recycler.

Figure 4 shows pictures taken of the 3rd batch injection for three different Recycler cycles. The first picture shows that the 2nd batch was injected at the correct position and the 3rd batch was injected early by one bucket. The last bunch on the 2nd batch was kicked out by the injection kicker for the 3rd batch. The second picture shows that the 2nd batch was injected earlier by one bucket and the 3rd batch was injected in the correct position. The last picture shows where the batch was supposed to be injected.

## BEAM STUDY AND OPERATION

Figure 5 illustrates 3 phase lock modes used for the jitter compensation. Each mode has a different divide-by-32 final phase resulting in a different number of Booster to Recycler rf phase wraps, but the same un-divided Booster rf to Recycler rf final phase offset. The difference seen when selecting different divide-by-32 final phase offsets is that the phase lock will be completed in more or less phase wraps of the Booster rf slipping past the Recycler rf. By choosing different divide-by-32 final phase offsets, we can control the number of rf periods, or buckets, that the Notch advances during the phase lock process. When the slow phase lock process takes control of the Booster rf frequency source, the acceleration phase lock control signal is frozen, the radial position control which normally adjusts rf phase to control beam position is disabled and the cogging feedback is frozen. For the last 3 msec the phase lock has sole control.

When the phase lock process starts, it signals the cogging controls. The cogging controls can predict at that time whether one more or one less wrap of the Booster to Recycler phase is required and signals the phase lock system with which divide-by-32 final phase mode to execute.

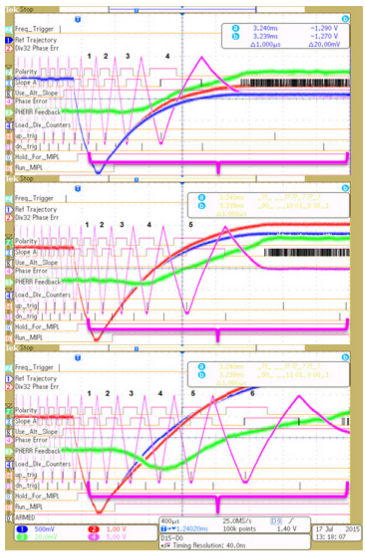

Figure 5: Illustration of the 3 phase lock modes.

Figure 6 shows the WCM signal that was measured after the 4th batch injection with the scope triggered 200 times. Each trace in this mountain range plot represents the bunch signal for a Recycler cycle. The middle picture in Figure 6 shows that the gap between the 2nd batch and the 3rd batch which has 3 empty buckets every pulse when the jitter compensation was working. With the jitter compensation removed, we see in the right-hand picture, that there were bunches shifted into the gap. It was found that beam loss at the Recycler injection was reduced by 20% with the jitter compensation.

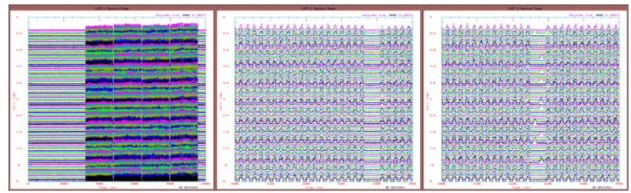

Figure 6: WCM signals in the Recycler. Each trace was triggered after 4th batch injection for 200 Recycler cycles.

Figure 7 shows the Beam intensity signal in the Booster (green) and in the Recycler (Red). The yellow traces are integrated beam loss at Recycler injection. The left-hand plot is with the jitter compensation and the right, without. The loss was increased at random injection times without jitter compensation and it was reduced by 20% with it.

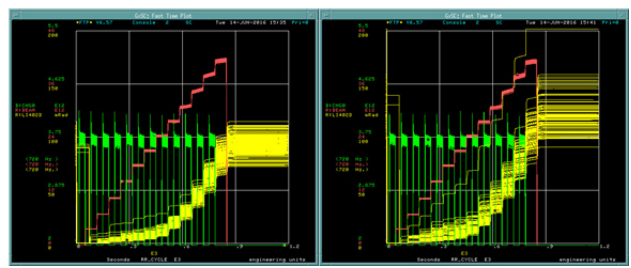

Figure 7: Beam intensity in the Booster (green) and Recycler (red), and an integration of the beam loss at Recycler injection (yellow).

## CONCLUSION

For the Fermi Accelerator complex to provide higher intensity beams, beam losses in all parts of the machine must be addressed. By implementing two digital mode selection signals between the cogging system and the phase lock controller, the bucket jitter was corrected. The loss at recycler injection was reduced by 20%.